\documentclass[reqno]{amsart}
\usepackage{amsmath,amsthm,amssymb}

\newcommand{\cprime}{\/{\mathsurround=0pt$'$}}
\newcommand{\F}{\mathcal{F}}
\newcommand{\E}{\mathcal{E}}
\newcommand{\FE}{\mathcal{F_\mathcal{E}}}
\newcommand{\I}{\mathcal{I}}
\newcommand{\Id}{\mathbf{I}}

\newcommand{\al}{\alpha}

\newcommand{\ch}{\mathrm{char}\,}
\newcommand{\la}{\lambda}
\newcommand{\tr}{\mathrm{tr}\,}

\newtheorem{theorem}{Theorem}

\newtheorem{proposition}{Proposition}
\newtheorem*{i_theorem}{Theorem}

\theoremstyle{definition}

\theoremstyle{remark}
\newtheorem{remark}{Remark}

\begin{document}
\title[Conservation laws and related linear algebra problems]
{Conservation  laws for multidimensional systems \\ 
and related linear algebra problems}
\author[Sergei Igonin]{Sergei Igonin}
\address{Independent University of Moscow; 
University of Twente, the Netherlands}
%\address{University of Twente, Faculty of Mathematical Sciences,
%P.O. Box 217, 7500 AE Enschede, the Netherlands}
\email{igonin@mccme.ru}

\begin{abstract}
We consider multidimensional systems of PDEs of generalized evolution 
form with $t$-derivatives of arbitrary order on the left-hand
side and with the right-hand side dependent on lower order 
$t$-derivatives and arbitrary space derivatives.
For such systems 
we find an explicit necessary condition for existence of higher 
conservation laws in terms of the system's symbol.
For systems that violate this condition we give an effective upper
bound on the order of conservation laws. 
Using this result, we completely describe 
conservation laws for viscous transonic equations, 
for the Brusselator model, and the Belousov-Zhabotinskii system.  
To achieve this, we 
solve over an arbitrary field 
the matrix equations $SA=A^tS$ and $SA=-A^tS$ 
for a quadratic matrix $A$ and its transpose $A^t$, 
which may be of independent interest. 

\end{abstract}
\maketitle

\section{Introduction}

It is well known that conservation laws are of fundamental importance
for clarifying the structure of PDEs. In particular, 
a common feature of 
soliton equations is to have conservation laws of arbitrarily high
order. Existence of higher order conservation laws
imposes very strong conditions on a system of PDEs. 
Explicit formulation of these conditions would help to classify 
integrable systems of a given type. 

The straightforward study of the conserved current condition is
hampered by the fact that one is interested in equivalence classes 
of conserved currents modulo trivial ones. 
Therefore, it is convenient to switch from a conserved current to its
characteristic, which is the same for
equivalent currents and satisfies the equation adjoint to the
linearization of the initial system \cite{rb,grif,olver,vin}. 

Thus a part of the problem is to determine conditions for the adjoint 
linearized equation to have higher order solutions $\chi$.
In the present article we perform the first natural step in this
direction. For determined, possibly multidimensional, systems of PDEs
we find the conditions imposed on the symbol of the system by
the fact that some higher order vector-functions satisfy 
the adjoint linearized equation modulo lower order terms.

More precisely, we consider systems of generalized evolution form
\begin{gather}
  \label{i_sys}
  \frac{\partial^h u^i}{\partial t^h}=F^i(x_j,t,\frac{\partial^s u^a}{\partial t^s},
  \frac{\partial^{i_1+\dots+i_n} u^b}{\partial x_1^{i_1}\dots\partial x_n^{i_n}}),\\ 
   i,\,a,\,b=1,\dots,m,\, \ \ j=1,\dots,n,\,\ \ s=0,\dots,h-1,
   \ \ i_1+\dots+i_n\le N,  \notag
\end{gather}
with $t$-derivatives of fixed order $h>0$ on the left-hand
side and with the right-hand side dependent on lower order 
$t$-derivatives and arbitrary space derivatives.

To any $m$-component vector-function $\chi$ of the variables
$x_j,\,t,\,u^i$ and their derivatives 
we associate its symbol with respect to the space variables $x_j$, which is 
an $m\times m$ matrix, whose entries are homogeneous polynomials
in $n$ variables of degree equal to the order $\mathrm{o}(\chi)$ of $\chi$
with respect to $x_j$.

Let $A$ be the symbol of the right-hand side of \eqref{i_sys}
and $S$ be the symbol of the characteristic $\chi$ of a
conservation law for \eqref{i_sys}.
It turns out that if $\mathrm{o}(\chi)>O$, where $O\le N$ is some constant 
associated to \eqref{i_sys}, 
then the adjoint linearized equation implies the matrix equation
\begin{equation}
  \label{i_first}
  SA=(-1)^{N+h}A^tS.
\end{equation}
Here and below $A^t$ is the transpose of $A$.

A linear algebra problem arises naturally: for what matrices $A$ does there
exist a nonzero matrix $S$ such that \eqref{i_first} holds? In
addition, since for known integrable systems there are normally 
higher conservation laws with nonsingular $S$, 
one is also interested for which $A$ the matrix $S$ can be taken
nonsingular, i.e., when the matrices $A$ and $\pm A^t$ are similar 
(conjugate). 

In solving these problems there is a difference between the cases $n=1$
and $n>1$. If $n=1$, one can switch 
from homogeneous polynomial in one variable 
matrices $A$ and $S$ to the corresponding matrices of
coefficients and, allowing the coefficients to be complex, make use of
the Jordan normal form \cite{foltinek}. While if $n>1$ then the entries of
the matrices belong to the field of rational functions in several
variables, which is essentially not algebraicly closed, hence the
Jordan normal form is not generally applicable.
Using more sophisticated algebraic technique, 
we prove the following effective criteria. 
\begin{i_theorem}
For any $m\times m$ 
matrix $A$ with the entries from an arbitrary field $F$ and the characteristic
polynomial $d(\lambda)\in F[\la]$ we have the following.
\begin{enumerate}
\item The matrices $A$ and $A^t$ are always similar.
\item A nonzero $m\times m$ matrix $S$ such that $SA=-A^tS$ exists if
  and only if the polynomials $d(\la)$ and $d(-\la)$ have a common
  divisor of positive degree. 
\item The matrices $A$ and $-A^t$ are similar if and only if all the
  invariant factors $d_i(\la)$ of $A$ \textup{(}certain divisors of the
  characteristic polynomial \cite{kurosh}\textup{)} satisfy 
  $d_i(-\la)=(-1)^{\deg d_i(\la)} d_i(\la)$. 
  In particular, $d(\la)=(-1)^{\deg d(\la)} d(\la)$, which in the case
  $\ch F\neq 2$ implies $\tr A=0$ and, if $m$ is odd, $\det A=0$.
\end{enumerate}  
\end{i_theorem}
Statements $2$ and $3$ of the theorem 
give a necessary condition for existence of
higher conservation laws for systems \eqref{i_sys} with odd $N+h$.
In particular, a scalar equation ($m=1$) of the form \eqref{i_sys}
with odd $N+h$ can not have conservation laws of order greater than $O$.  
For different ways to write system \eqref{i_sys} in the generalized evolution
form the symbols $A$ and the resulting conditions are generally
different. In order to have higher conservation laws a system of PDEs 
must satisfy all conditions obtained from various ways to write it 
in the generalized evolution form.

Let us discuss the previous research on this theme. 
It seems that only evolution systems ($h=1$) in one space variable
($n=1$) were studied in this respect. For such systems equation 
\eqref{i_first} was obtained by a similar technique in
\cite{mikh} and rediscovered in \cite{foltinek}.
In \cite{mikh} it is noticed that $SA=-A^tS$ for nonsingular $S$
implies $\det(A+\la\Id)=(-1)^m\det(A-\la\Id)$, 
which is a weaker version of our Statement $3$.
Here and below $\Id$ is the unity matrix.
In \cite{foltinek} Statement $2$ is proved for complex matrices, 
and the corresponding necessary
condition for existence of conservation laws of order greater than the
order of the evolution system is formulated. Even in this 
simplest case our result is stronger, since 
the upper bound $O$ is normally much smaller than the
order of the system (see the examples in Section \ref{ex}).  

The paper is organized as follows.
In Section \ref{ccl} the method of characteristics of 
conservation laws is recalled. 
We specify the method for systems of generalized evolution form 
in Section \ref{gef} and derive equation \eqref{i_first} in Section
\ref{main}. The above theorem on quadratic matrices is proved 
in Section \ref{lap}. In Section \ref{n_c} we explicitly
formulate the obtained necessary conditions for existence of higher
conservation laws. Finally, Section \ref{ex} 
contains some mathematical physics equations of the form \eqref{i_sys}, 
which violate these conditions and, therefore, do not have
conservation laws of order greater than $O$. 
This result allows us to describe
all conservation laws for two basic equations 
in the theory of viscous transonic gas flows 
(see, for example, \cite{larkin,vis,ryzhov,vel} and references therein)
and for two popular reaction-diffusion systems:
the Brusselator model and the Belousov-Zhabotinskii system
\cite[Section 15.4]{pdes}.

\section{Characteristics of conservation laws}
\label{ccl}

This is a brief review of the method of characteristic for computation
of conservation laws. We refer to \cite{rb,grif,olver,vin} for more
details.

Consider a system $\E$ of differential equations 
\begin{equation}
  \label{sys0}
  F_s(y_i,v^j,\dots,v^k_I,\dots)=0,\ s=1,\dots,p,   
\end{equation}
with independent variables $y_1,\dots,y_a$,
unknown functions $v^1,\dots,v^b$, and 
$$
v^k_I=\frac{\partial^{|I|}v^k}{\partial y_1^{i_1}\dots\partial y_a^{i_a}},\ \ 
I=(i_1,\dots,i_a)\in\mathbb{Z}_+^a,
$$
being their derivatives. Here and below $\mathbb{Z}_+$ is the set of
nonnegative integers and $|I|=i_1+\dots+i_a$.
 
Let $\F$ be the algebra of smooth functions 
of the variables $y_i$, $u^j$, and $u^j_I$. Although the whole set of
the variables is infinite, each function is supposed to depend 
only on a finite subset. Denote by $\F_\E$ the
quotient algebra with respect to the ideal $\I$ generated by the left-hand
sides of equations \eqref{sys0} and their differential consequences
$D_{y_{i_1}}\dots D_{y_{i_k}}(F_s)\in\F$. Here 
\begin{equation*}
\label{dy}
D_{y_i}=\frac{\partial}{\partial y_i}+\sum_{j,I}u^j_{I+1_i}\frac{\partial}{\partial u^j_{I}}
\end{equation*}
is the \emph{total derivative with respect to} $y_i$, where 
$1_i$ is the multi-index with $1$ at the $i$-th place, the other indices
of $1_i$ being zero.
For two equivalent functions $f_1,\,f_2\in\F,\,f_1-f_2\in\I,$ one has 
$f_1(y_i,v^j(y_i))=f_2(y_i,v^j(y_i))$ for any local solution
$v^j(y_i)$ to \eqref{sys0}.

By definition, 
the ideal $\I$ is invariant under the action of $D_{y_i}$, 
which, therefore, defines a derivation $\bar D_{y_i}$ 
of $\F_\E$. A \emph{conserved current for} \eqref{sys0} is an $a$-tuple 
$J=(J_1,\dots,J_a)$, where $J_k\in\FE$, that satisfies the equation
\begin{equation}
\label{cc}
\sum_{i=1}^a\bar D_{y_i}(J_i)=0.
\end{equation}
A conserved current is called \emph{trivial}, if it has the form
\begin{equation*}
J_k=\sum_{l<k} \bar D_l(\mathcal{L}_{lk})-
\sum_{k<l} \bar D_l(\mathcal{L}_{kl})
\end{equation*}
for some functions $\mathcal{L}_{kl}\in\FE,\,1\le k<l\le a$.
Two conserved currents are said to be \emph{equivalent} if they differ
by a trivial one. \emph{Conservation laws} are defined to be the
equivalent classes of conserved currents.

Let $\tilde J_k\in\F$ be such that $J_k=\tilde J_k+\I$.
Identity \eqref{cc} means that
\begin{equation}
\label{F}
\sum_{i=1}^aD_{y_i}(\tilde J_i)=\sum_{s,I}g^s_ID_y^I(F_s)
\end{equation}
for some functions $g^s_I\in\F$, only a finite number of which is
nonzero. Here and in what follows for each multi-index
$I=(i_1,\dots,i_a)$ we denote 
$D_y^I=D_{y_1}^{i_1}\dots D_{y_a}^{i_a}$. Consider the functions
\begin{equation}
\label{gen_chi}
\tilde\chi_s=\sum_{I}(-1)^{|I|}D_y^I(g^s_I),\ \ s=1,\dots,p.
\end{equation}
Generally speaking, representation \eqref{F} and functions 
$\tilde\chi_s$ are not uniquely defined by the conserved current $J$.
Assume that system \eqref{sys0} is non-overdetermined and nondegenerated
\cite{rb,olver}, then the corresponding elements 
$\chi_s=\tilde\chi_s+\I$ of $\FE$ 
are well-defined by $J$ and are all zero if and only if $J$ is trivial. 
The $p$-tuple $\chi=(\chi_1,\dots,\chi_p)$ is the same for equivalent
conserved currents and is called the \emph{characteristic} (or
\emph{generating function} \cite{rb,vin}) of the 
corresponding conservation law. 

In addition, $\chi$ satisfies the \emph{adjoint linearized equation} 
\begin{equation}
\label{ale}
K(\chi)=0,
\end{equation}
where $K$ is the $b\times p$ matrix differential operator with the
entries 
\begin{equation}
\label{alo}
[K]_{ij}=\sum_I 
  (-1)^{|I|}\bar D_y^I\circ\frac{\partial F_j}{\partial v^i_{I}}.
\end{equation}

The homological interpretation of these concepts can be found in
\cite{rb,grif,vin}.

\section{Formulas for systems of generalized evolution form}
\label{gef}

Consider a system $\E$ of $m$ partial differential equations in $n+1$
independent variables $t,x_1,\dots,x_n$ and $m$ unknown functions
$u^1,\dots,u^m$ of the form 
\begin{equation}
  \label{sys}
  \frac{\partial^h u^i}{\partial t^h}=F^i,\ \ i=1,\dots,m,
\end{equation}
where $F^i$ are smooth functions of the variables $x_j,t,u^a$ and
the following derivatives 
\begin{equation}
  \label{depend}
  \frac{\partial^{i_1+\dots+i_n} u^b}{\partial x_1^{i_1}\dots\partial x_n^{i_n}},\,
  i_1,\dots,i_n\in\mathbb{Z}_+,\ \ \
  \frac{\partial^s u^a}{\partial t^s},\,s<h.
\end{equation}
For each multi-index $I=(i_1,\dots,i_n)\in\mathbb{Z}_+^n$
and integer $s\ge 0$ we denote 
\begin{equation}
  \label{u}
  u_{s,I}^i=\frac{\partial^{|I|+s} u^i}{\partial t^s\partial x_1^{i_1}\dots
  \partial x_n^{i_n}}. 
\end{equation}
Let us describe the algebra $\FE$ for this system.
By induction on $s$, it follows from \eqref{sys} that
the derivatives $u_{s,I}^i$ with $s\ge h$ are expressed in terms of 
\begin{equation}
\label{coor}
t,\ \ x_j,\ \ \ u_{s,I}^i,\,s<h.  
\end{equation}
Therefore,
for each function of $t,\,x_j,\,u^i$ and arbitrary derivatives
\eqref{u} there is a unique equivalent modulo \eqref{sys} function
of variables \eqref{coor}.
Thus we can identify $\FE$ with the algebra of smooth functions 
of variables \eqref{coor}. Below all functions are
supposed to be from $\FE$. 

The restrictions $\bar D_{x_i},\bar D_t\colon\FE\to\FE$ 
of the total derivatives are written in
coordinates \eqref{coor} as follows
\begin{gather}
   \label{dx}
   \bar D_{x_i}=\frac{\partial}{\partial x_i}+\sum_{k,I,s}u^k_{s,I+1_i}\frac{\partial}{\partial u^k_{s,I}},\notag\\
   \label{dt}
   \bar D_t=\frac{\partial}{\partial t}+\sum_{k,I,\,s<h-1}u^k_{s+1,I}\frac{\partial}{\partial u^k_{s,I}}+
   \sum_{k,I}D_x^I(F^k)\frac{\partial}{\partial u^k_{h-1,I}}. 
\end{gather}
The equation $\bar D_tJ_0+\sum_{i=1}^n\bar D_{x_i}J_i=0$ for a conserved 
current $J=(J_0,J_1,\dots,J_{n})$ implies the identity
\begin{equation*}
  D_tJ_0+\sum_{i=1}^n D_{x_i}J_i=\sum_{i,I}\frac{\partial J_0}{\partial u^i_{h-1,I}}D_x^I(u^i_{h,0}-F^i),
\end{equation*}
which is the specification of \eqref{F} for system \eqref{sys}.
According to general formula \eqref{gen_chi}, the characteristic 
$\chi=(\chi_1,\dots,\chi_m)$ is computed as follows
\begin{equation}
  \label{chi}
  \chi_i=\sum_I(-1)^{|I|}\bar D_x^I\bigl(\frac{\partial J_0}{\partial u^i_{h-1,I}}\bigl).
\end{equation}

From \eqref{ale} and \eqref{alo} we see that 
the characteristic regarded as a column vector satisfies the equation
\begin{equation}
  \label{eq_xi}
  (-1)^h\bar D_t^h(\chi)=L(\chi),
\end{equation}
where $L$ is the $m\times m$ matrix differential
operator with the entries 
\begin{equation}
  \label{lin}
  [L]_{ij}=\sum_I 
  (-1)^{|I|}\bar D_x^I\circ\frac{\partial F^j}{\partial u^i_{0,I}}+
  \sum_{s=0}^{h-1}(-1)^s\bar D_t^s\circ\frac{\partial F^j}{\partial u^i_{s,0}}.
\end{equation}

\section{Solving the adjoint equation for the highest order terms}
\label{main}

For a (vector-)function $f$ the maximal integer $k$ such that 
${\partial f}/{\partial u_{s,I}^i}\neq 0$ for some 
$0\le s< h,\,1\le i\le m,\,|I|=k$ is
called the {\it order} 
of $f$ and denoted by $\mathrm{o}(f)$. If ${\partial f}/{\partial u_{s,I}^i}=0$ for all 
$s,\,i,\,I$ then we set $\mathrm{o}(f)=-1$.
The maximal integer $s<h$ such that ${\partial f}/{\partial u_{s,I}^i}\neq 0$ for
some $1\le i\le m,\,|I|=\mathrm{o}(f)$ is denoted by $\mathrm{t}(f)$.
The order of the characteristic of a conservation law for \eqref{sys} 
is called the \emph{order} of the conservation law.

Consider the ring $\F[q_1,\dots,q_n]$
of polynomials in $n$ variables with $\F$ as the ring of coefficients.
For each multi-index $I=(i_1,\dots,i_n)$ denote by $q^I$ the monomial 
$q_1^{i_1}\dots q_n^{i_n}\in \F[q_1,\dots,q_n]$.    
For any $k$-component vector-function 
$\chi=(\chi_1,\dots,\chi_k)$ and two integers $a,\,b\ge 0$ let 
${\mathbf{S}}^{a,b}(\chi)$ be the $k\times m$ matrix with the entries
$$
[{\mathbf{S}}^{a,b}(\chi)]_{ij}=\sum_{|I|=a}\frac{\partial \chi_i}{\partial u^j_{b,I}}q^I\in 
\F[q_1,\dots,q_n].
$$
We call the nonzero matrix ${\mathbf{S}}^{\mathrm{o}(\chi),\mathrm{t}(\chi)}(\chi)$ the 
\emph{symbol} of $\chi$ and denote it by ${\mathbf{S}}_\chi$.

Let $A$ be the symbol of the right-hand side $(F^1,\dots,F^m)$ of
\eqref{sys} and set $N=\mathrm{o}(F^1,\dots,F^m)>0$. By assumption \eqref{depend},
one has $\mathrm{t}(F^1,\dots,F^m)=0$. Therefore, by definition, 
\begin{equation}
  \label{A}
  [A]_{ij}=\sum_{|I|=N}\frac{\partial F^i}{\partial u^j_{0,I}}q^I.
\end{equation}

Applying the Leibniz rule, differential operators \eqref{lin} 
can be uniquely rewritten in the usual form
\begin{equation*}
  [L]_{ij}=\sum_{|I|\le N} f_I^{ij}\bar D_x^I+ 
  \sum_{s<h}g_s^{ij}\bar D_t^s.
\end{equation*}
In particular, from definition \eqref{lin} one has
\begin{equation}
  \label{fI}
  f_I^{ij}=(-1)^N\frac{\partial F^j}{\partial u^i_{0,I}}\ \ \forall\,I:|I|=N.
\end{equation}
We set 
\begin{equation}
  \label{O}
  O=\max_{i,j,I,s}\{-1,\,\mathrm{o}(f_I^{ij})-N,\,\mathrm{o}(g_s^{ij})-N\}.
\end{equation}
From definition \eqref{lin} it follows that $O\le N$.

\begin{theorem}
\label{saas}
Let $\chi$ be the characteristic of a conservation law for \eqref{sys}.
If $\mathrm{o}(\chi)>O$ then we have 
\begin{equation}
  \label{cond}
  {\mathbf{S}}_{\chi} A=(-1)^{N+h}A^t{\mathbf{S}}_{\chi}.
\end{equation}
\end{theorem}
\begin{proof}
Set $a=N+\mathrm{o}(\chi),\,b=\mathrm{t}(\chi)<h$.
Equation \eqref{eq_xi} implies, in particular,
\begin{equation}
  \label{S}
  (-1)^h{\mathbf{S}}^{a,b}({\bar D_t^h(\chi)})={\mathbf{S}}^{a,b}({L(\chi)}).
\end{equation}

Set
$$
\Delta=\sum_{k,I,\,s<h-1}u^k_{s+1,I}\frac{\partial}{\partial u^k_{s,I}}.
$$
Clearly, for any vector-function $\psi$ with $\mathrm{o}(\psi)\ge 0$ and  $\mathrm{t}(\psi)<h-1$ we have
\begin{equation}
\label{Delta} 
\mathrm{o}(\Delta(\psi))=\mathrm{o}(\psi),\quad 
\mathrm{t}(\Delta(\psi))=\mathrm{t}(\psi)+1,\quad   
{\mathbf{S}}_{\Delta(\psi)}={\mathbf{S}}_{\psi}.
\end{equation}

By formula \eqref{dt} and assumption \eqref{depend}, 
for any vector-function $\chi$ with $\mathrm{o}(\chi)\ge 0$ the 
vector-function $\bar D_t^h(\chi)$ does not depend on the coordinates 
$u_{s,I}^j$ with $|I|>a$ or $|I|=a,\,s>b$. 
Moreover, from the whole expression 
$$
\bar D_t^h(\chi)=
\Bigl(\frac{\partial}{\partial t}+\Delta+
\sum_{k,I}D_x^I(F^k)\frac{\partial}{\partial u^k_{h-1,I}}\Bigl)^h(\chi) 
$$
only the part 
\begin{equation*}
  \label{part}
\Delta^b\circ
\Bigl(\sum_{k,I}D_x^I(F^k)\frac{\partial}{\partial u^k_{h-1,I}}\Bigl)\circ
\Delta^{h-1-b}(\chi) 
\end{equation*}
contributes to ${\mathbf{S}}^{a,b}({\bar D_t^h(\chi)})$. 

Therefore, by the definition of ${\mathbf{S}}^{a,b}$, properties \eqref{Delta},  
and formula \eqref{A}, we obtain 
\begin{multline}
  \label{St}
  {\mathbf{S}}^{a,b}(\bar D_t^h(\chi))={\mathbf{S}}^{a,b}\Bigl(\Delta^b\circ
  \Bigl(\sum_{k,I}D_x^I(F^k)\frac{\partial}{\partial u^k_{h-1,I}}\Bigl)\circ
  \Delta^{h-1-b}(\chi)\Bigl)=\\
  ={\mathbf{S}}^{a,0}
  \Bigl(\sum_{k,|I|=\mathrm{o}(\chi)}
  D_x^I(F^k)\frac{\partial\chi}{\partial u^k_{b,I}}\Bigl)=
  {\mathbf{S}}_\chi A.
\end{multline}

Now let us compute the right-hand side of \eqref{S}.
In the case $\mathrm{o}(\chi)>O$ 
only the part $\sum_{|I|=N} f_I^{ij}\bar D_x^I$ of $[L]_{ij}$ contributes to
${\mathbf{S}}^{a,b}({L(\chi)})$, since $\bar D^s_t(\chi)$ for $s<h$ does not depend
on $u_{p,I}^j$ with $|I|=a,\,p\ge b$ and the number $a$ is greater
than the order of any coefficient $f_I^{ij}$ or $g_s^{ij}$. 
Therefore, taking into account formulas \eqref{fI} and \eqref{A}, 
we obtain 
\begin{equation*}
  \label{SL}
 {\mathbf{S}}^{a,b}({L(\chi)})=(-1)^NA^t{\mathbf{S}}_\chi.
\end{equation*}
Combining this with \eqref{St} and \eqref{S}, one gets \eqref{cond}.
\end{proof}
In the next section 
we study the conditions imposed on $A$ by equation \eqref{cond}. 

\section{Linear algebra problems}
\label{lap}

For a ring $R$ we denote by $M_k(R)$ the ring of 
$k\times k$ matrices with entries from $R$.
Consider an arbitrary field $F$ and denote by $\tilde F$ its 
algebraic closure, i.e., the minimal algebraicly closed extention of
$F$.

\begin{theorem}
\label{la1}
Let $A\in M_k(F)$ and let $d(\la)=\det(A-\la\Id)$ be the characteristic
polynomial of $A$. A nonzero matrix $S\in M_k(F)$ such that 
\begin{equation}
  \label{AS}
  SA=-A^tS
\end{equation}
exists if and only if the polynomials 
$d(\la)$ and $d(-\la)$ have a common divisor of
positive degree. Equivalently, 
there are two roots $\la_1,\,\la_2\in\tilde F$ of $d(\la)$ 
such that $\la_1+\la_2=0$.
\end{theorem}
\begin{proof}
The polynomials $d(\la)$ and $d(-\la)$ have a common divisor of
positive degree if and only if they have a common root $\la$ 
in $\tilde F$, i.e., both $\la$ and $-\la$ are roots of $d(\la)$.

If \eqref{AS} holds then for any similar to $A$ matrix
$A'=CAC^{-1},\,C\in M_k(F)$, one has $S'A'=-{A'}^tS'$ 
with $S'={C^{-1}}^tSC^{-1}$. 
Let us regard $A$ as a matrix from $M_k(\tilde F)\supset M_k(F)$.
Then we can assume $A$ to be in the Jordan normal form. 
For such $A$ one can easily show that the linear map
\begin{equation*}
M_k(\tilde F)\to M_k(\tilde F),\ S\mapsto SA+A^tS,
\end{equation*}
has a nontrivial kernel if and only if there are 
two eigenvalues $\la_1,\,\la_2\in\tilde F$ of $A$ 
such that $\la_1+\la_2=0$ (see \cite{foltinek} for $F=\mathbb{C}$).

It remains to prove that if \eqref{AS} holds for some
$S\in M_k(\tilde F)$ then there is nonzero $S'\in M_k(F)$ such that
$S'A=-A^tS'$. Consider a (possibly infinite) 
basis $\{a_i\}$ of $\tilde F$ regarded as a vector space over $F$.
One has $S=\sum_ia_iS_i$, where $S_i\in M_k(F)$. 
Since $A\in M_k(F)$, equation \eqref{AS} implies $S_iA=-A^tS_i$.
\end{proof}

Recall some criteria for two matrices $A,\,B\in M_k(F)$ 
to be similar (see, for example, \cite[Chapter 13]{kurosh}). 
Consider the ring $F[\la]$ of polynomials in one
variable. A matrix $C\in M_k(F[\la])$ is said to be \emph{unimodular}
if $\det C$ is nonzero and belongs to $F$. For any matrix $A\in M_k(F)$
the matrix $A-\la\Id\in M_k(F[\la])$ admits a \emph{canonical decomposition} 
\begin{equation}
  \label{dec}
  A-\la\Id=C_1DC_2,\ \ C_1,\,D,\,C_2\in M_k(F[\la]),
\end{equation}
such that $C_1,\,C_2$ are unimodular, while $D$ is diagonal.
Moreover, the polynomials $d_i=[D]_{ii}$ have the leading coefficient $1$, and   
$d_{i+1}$ is divisible by $d_{i}$ for each $i=1,\dots,k-1$.
Then the $k$ polynomials $d_i$ are defined uniquely by $A$ and are called the
\emph{invariant factors} of $A$. Note that there is a simple procedure
to compute the invariant factors \cite[Chapter 13]{kurosh}.

\begin{proposition}[{\cite[Chapter 13]{kurosh}}]
\label{invfac}
Two matrices $A,\,B\in M_k(F)$ are similar if and only if they have
the same invariant factors.
\end{proposition}

We call a polynomial $d(\lambda)=\sum_ia_i\lambda^i\in F[\la]$  
\emph{skew} if $d(-\lambda)=(-1)^{\deg d(\la)}d(\lambda)$, i.e.,
for all $i\equiv \deg d(\la)+1\mod 2$ one has $a_i=0$.

\begin{theorem}
\label{la2}
For any $A\in M_k(F)$ we have the following. 
\begin{enumerate}
\item 
\label{A+}
The matrices $A$ and $A^t$ are similar.
\item 
\label{A-}
The matrices $A$ and $-A^t$ are similar if and only if 
  each invariant factor of $A$ is skew. In this case the
  characteristic polynomial is also skew. In particular, in the case
  $\ch F\neq 2$ we have $\tr A=0$ and, if $k$ is odd, $\det A=0$.
\end{enumerate}
\end{theorem}
\begin{remark}
Note that in the case $\ch F=2$ the second statement of 
this theorem as well as Theorem~\ref{la1} are trivial.
\end{remark}
\begin{proof}
Consider canonical decomposition \eqref{dec} for $A$. 
Taking the transpose, we obtain 
\begin{equation}
  \label{dec1}
  A^t-\la\Id={C_2}^tD{C_1}^t, 
\end{equation}
which is a canonical decomposition for $A^t$, since
${C_1}^t,\,{C_2}^t$ are clearly unimodular. Therefore, the invariant
factors of $A^t$ are the same, 
which, by Proposition~\ref{invfac}, implies that $A$ and $A^t$ are similar.

Multiplying \eqref{dec1} by $-1$ and substituting $-\la$ in place of
$\la$, we obtain
\begin{equation}
  \label{dec2}
  -A^t-\la\Id=-{C_2}^t(-\la)D(-\la){C_1}^t(-\la). 
\end{equation}
Denote $C_1'=-{C_2}^t(-\la)$, $C_2'=T{C_1}^t(-\la)$, and $D'=D(-\la)T$,
where $T\in M_k(F)$ is the diagonal matrix with the entries 
$[T]_{ii}=(-1)^{\deg[D]_{ii}}$. From \eqref{dec2} we obtain the
canonical decomposition $-A^t-\la\Id=C_1'D'C_2'$ for $-A^t$. 
According to Proposition~\ref{invfac}, 
$A$ and $-A^t$ are similar if and only if $D'=D$,
which says that all the invariant factors $[D]_{ii}$ of $A$ 
are skew. In this case the characteristic polynomial is also skew, 
since from \eqref{dec} it is evidently equal to the product 
of the invariant factors multiplied by $(-1)^k$.
\end{proof}

\section{Necessary conditions for existence of higher conservation laws}
\label{n_c}

According to Theorem~\ref{saas}, 
a necessary condition for existence of conservation laws for
\eqref{sys} of order greater than $O$ is that there is a 
nonzero $m\times m$
matrix ${\mathbf{S}}_\chi$ with entries from $\F[q_1,\dots,q_n]$ such that 
\eqref{cond} holds.
Let us treat $A$ and ${\mathbf{S}}_\chi$ as matrices with entries from the
field $F$ of 
rational functions in $n$ variables $q_1,\dots,q_n$.
Then Theorem~\ref{la1} implies the following.
\begin{theorem}
\label{nc}
If $N+h$ is odd and system \eqref{sys} possesses a conservation law of order
greater than $O$, then the characteristic polynomial 
$d(\la)=\det(A-\la\Id)$ and the polynomial $d(-\la)$ have a common
divisor of positive degree. Equivalently, 
there are eigenvalues $\la_1,\,\la_2$ of $A$ 
\textup{(}possibly in the algebraic closure of $F$\textup{)} 
such that $\la_1+\la_2=0$.   
\end{theorem}
\begin{remark}
Evidently, introducing the new dependent variables 
\[
u^{i,s}=\frac{\partial^s u^i}{\partial t^s},\ \ i=1,\dots,m,\ s=0,\dots,h-1,
\]
we can rewrite \eqref{sys} in the usual evolution form.
But if $h>1$ then the symbol of the right-hand side of the obtained
evolution system has zero determinant 
and, therefore, automatically satisfies the condition in
Theorem \ref{nc}, even if the initial system does not meet this
condition. Therefore, it is essential to consider the generalized
evolution form.
\end{remark}

Analyzing examples of known soliton equations, 
we can conjecture that for \eqref{sys} to be integrable 
there must exist 
higher order conservation laws with nonsingular matrix ${\mathbf{S}}_\chi$.
Therefore, it is worth formulating 
a necessary condition for existence of such \emph{nonsingular} 
conservation laws. According to Theorem~\ref{la2} (\ref{A-}), 
we obtain the following criterion. 
\begin{theorem}
\label{maximal}
If $N+h$ is odd and system \eqref{sys} has a nonsingular 
conservation law of order greater than $O$ then 
all the invariant factors of the matrix $A$ are skew.
In this case its characteristic polynomial is also skew.
In particular, $\tr A=0$ and, if $m$ is odd, $\det A=0$.
\end{theorem}
If $m=1$ then $A$ is a nonzero $1\times 1$ matrix, and Theorem~\ref{nc}
implies the following.
\begin{theorem}
\label{scal}
A scalar equation 
\textup{(}$m=1$\textup{)} of the form \eqref{sys} with odd
$N+h$ can not have conservation laws of order greater than $O$.   
\end{theorem}
\begin{remark}
To obtain stronger conditions, it is sometimes useful to write a
system of PDEs in several ways in the form \eqref{sys}. 
For example, for a scalar evolution equation  
$u_{t}=u_{xxx}+u_{yy}+f(u,u_x,u_y,u_{xx})$ 
the condition is empty, since the sum of $h=1$ and $N=3$ is even. 
But rewriting the equation in the generalized evolution form 
with respect to $y$ as follows 
$u_{yy}=-u_{xxx}+u_{t}-f(u,u_x,u_y,u_{xx})$, 
we see, according to Theorem~\ref{scal}, 
that there are no higher conservation laws.
\end{remark}
\begin{remark}
If $N+h$ is even then, according to Theorem \ref{la2} (\ref{A+}),  
equation \eqref{cond} is always solvable and imposes no restrictions 
on the symbol $A$ of \eqref{sys}. That is, equation \eqref{eq_xi} 
is solvable with respect to the highest order terms. In this case,    
to obtain nontrivial conditions for \eqref{sys} to have 
higher conservation laws, 
deeper analysis of \eqref{eq_xi} is needed. 
\end{remark}

\section{Examples}
\label{ex}

\subsection{The viscous transonic equation}
The nonlinear viscous transonic equation
\begin{equation}
  \label{vistr}
  u_{tt}=-u_{xxx}+u_xu_{xx}-\frac{\al}{t}u_t
\end{equation}
describes the asymptotic form of a gas flow in the sonic region
(see \cite{vis,ryzhov} and references therein). 
The following conserved currents for \eqref{vistr} were
found in \cite{vis}
\begin{equation}
  \label{vis_cons}
  (u_tt^\al,\,u_{xx}t^\al-\frac{u_x^2}{2}t^\al),\ \ 
  (u_tt+(\al-1)u,\,u_{xx}t-\frac{u_x^2}{2}t). 
\end{equation}
All other conserved currents mentioned in \cite{vis} are trivial.
 
Let us show that \eqref{vis_cons} span the whole space of conservation
laws for \eqref{vistr}. We have
\begin{equation*}
  L=\bar D_x^3+\bar D_x^2\circ u_x -\bar D_x\circ u_{xx}+\bar D_t\circ\frac{\al}{t}=
  \bar D_x^3+u_x\bar D_x^2+u_{xx}\bar D_x+\frac{\al}{t}\bar D_t-\frac{\al}{t^2}.
\end{equation*}
According to \eqref{O}, one has $O=-1$. By Theorem~\ref{scal}, 
since $N+h=5$ is odd, the characteristic $\chi$  
of any conservation law is a function of $x,\,t$ only.
Equation \eqref{eq_xi} reads
\begin{equation*}
  \label{eqvis}
  \chi_{tt}=\chi_{xxx}+u_x\chi_{xx}+u_{xx}\chi_x+
  \frac{\al}{t}\chi_t-\frac{\al}{t^2}\chi.
\end{equation*}
This implies $\chi_x=0$ and
\begin{equation}
  \label{chi_t}
  \chi_{tt}=\frac{\al}{t}\chi_t-\frac{\al}{t^2}\chi.
\end{equation}
The characteristics $t^\al,\,t$ of conserved currents \eqref{vis_cons} span
the space of solutions to \eqref{chi_t}. Therefore, \eqref{vis_cons}
span the space of conservation laws.

\subsection{Another equation for viscous transonic flows}
The equation 
\begin{equation}
  \label{vistr1}
  v_{yy}=2v_{xt}+v_xv_{xx}-v_{zz}-\mu v_{xxx},
\end{equation}
where $\mu$ is a nonzero real constant,
models nonstationary transonic flows around a thin body with 
effects of viscosity and heat conductivity when the velocity of 
the gas is close to the local speed of sound, 
see \cite{larkin,vel} and references therein. 

For this equation we obtain
\begin{equation*}
  L=\mu\bar D_x^3+2\bar D_x\bar D_t-\bar D_z^2+v_x\bar D_x^2+v_{xx}\bar D_x.
\end{equation*}
By definition \eqref{O}, we have $O=-1$. According to Theorem~\ref{scal}, 
since $N+h=5$ is odd, the characteristic $\chi$  
of any conservation law is a function of $x,\,y,\,z,\,t$ only.
Equation \eqref{eq_xi} implies
\begin{equation}
  \label{chi_t1}
  \chi_x=0,\ \chi_{yy}+\chi_{zz}=0.
\end{equation}
Each function $\chi$ satisfying \eqref{chi_t1} is indeed the
characteristic of the conserved current
\begin{equation}
  \label{ccvis}
  \bar D_y(\chi v_y-\chi_yv)+\bar D_z(\chi v_z-\chi_z v)+\bar D_x(\mu\chi
  v_{xx}-\frac12\chi v_x^2-2\chi v_t)=0.
\end{equation}
Therefore, any conserved current for \eqref{vistr1} is equivalent to a
conserved current of the form \eqref{ccvis}.

\subsection{The Brusselator model} 
The Brusselator model governing certain chemical reactions is the
following multidimensional system
\begin{equation}
  \label{brus}
  \begin{aligned}
  v_t&=\sum_{i=1}^n \frac{\partial^2 v}{\partial x_i^2}+v^2w-(b+1)v+a,\\
  w_t&=c\sum_{i=1}^n \frac{\partial^2 w}{\partial x_i^2}-v^2w+bv,
  \end{aligned}
\end{equation}
where $a$, $b$, and $c\neq 0$ are real parameters \cite[Section 15.4]{pdes}. 
By definition \eqref{lin} we have
\begin{equation*}
  L=\left(
  \begin{array}{cc}
  \sum_i \bar D_{x_i}^2+2vw-(b+1)& -2vw+b\\
  v^2 & c\sum_i \bar D_{x_i}^2-v^2
  \end{array}\right).
\end{equation*}
The symbol $A$ is diagonal
with $[A]_{11}=\sum_iq_i^2$ and $[A]_{22}=c\sum_iq_i^2$.
By definition \eqref{O}, we have $O=-1$. By Theorem~\ref{nc}, 
since $[A]_{11}\neq 0$, 
$[A]_{22}\neq 0$, and $[A]_{11}+[A]_{22}\neq 0$, 
the characteristic $\chi=(\chi_1,\chi_2)$  
of any conservation law is a function of $x_i,\,t$ only.
For such $\chi$ equation \eqref{eq_xi} reads
\begin{equation}
  \label{chiBr}
  \begin{aligned}
    -\frac{\partial\chi_1}{\partial t}&=\sum_i \frac{\partial^2\chi_1}{\partial x_i^2}+
     (2vw-(b+1))\chi_1-(2vw-b)\chi_2,\\
    -\frac{\partial\chi_2}{\partial t}&=v^2\chi_1+c\sum_i \frac{\partial^2\chi_2}{\partial x_i^2}-v^2\chi_2.  
  \end{aligned} 
\end{equation}
Evidently, \eqref{chiBr} implies $\chi_1=\chi_2$. Then \eqref{chiBr} becomes 
\begin{equation}
  \label{chiBr1}
  \begin{aligned}
    -\frac{\partial\chi_1}{\partial t}&=\sum_i \frac{\partial^2\chi_1}{\partial x_i^2}-\chi_1,\\
    -\frac{\partial\chi_1}{\partial t}&=c\sum_i \frac{\partial^2\chi_1}{\partial x_i^2}.  
  \end{aligned} 
\end{equation}
Clearly, for $c=1$ the only solution to \eqref{chiBr1} is $\chi_1=0$, 
while for $c\neq 1$ the general solution is
\begin{equation}
  \label{chBr}
  \chi_1=\chi_2=G(x_i)\exp(\frac{ct}{c-1}),
\end{equation}
where $G(x_i)$ is an arbitrary solution to the equation
\begin{equation}
\label{G}
G+(c-1)\sum_i \frac{\partial^2G}{\partial x_i^2}=0.
\end{equation}
Each solution \eqref{chBr} is indeed the characteristic of the conserved
current 
\begin{multline*}
\bar D_t\bigl(G\exp(\frac{ct}{c-1})(v+w)\bigl)+\\
+\sum_{i=1}^n \bar D_{x_i}\Bigl(\exp(\frac{ct}{c-1})\bigl(
\frac{\partial G}{\partial x_i}(v+cw)-G(v_{x_i}+cw_{x_i})
-\frac{a}{n}\int G\,\mathrm{d}x_i\bigl)\Bigl)=0.
\end{multline*}
Thus if $c\neq 1$ then these conserved currents span the space of
conservation laws; while if $c=1$ then there are no nontrivial 
conservation laws for \eqref{brus}.

\subsection{The Belousov-Zhabotinskii system} 
This system describes certain chemical reactions and reads 
\cite[Section 15.4]{pdes}
\begin{equation*}
  \begin{aligned}
  v_t&=\sum_i \frac{\partial^2 v}{\partial x_i^2}+v(1-v-rw)+Lrw,\\
  w_t&=\sum_i \frac{\partial^2 w}{\partial x_i^2}-bvw-Mw.
  \end{aligned}
\end{equation*}
Here $r,\,L,\,b,\,M$ are real constants. 
Evidently, Theorem~\ref{nc} implies that this system does not possess
conservation laws of nonnegative order.
Similarly to the above examples, analysis of equation \eqref{eq_xi} for
characteristics of order $-1$ shows that 
in the nonlinear case $b\neq 0$ there are no nontrivial
conservation laws at all.

\section*{Acknowledgements}
The author is grateful to Iosif~Krasil{\cprime}shchik for his very
stimulating attention to this work and many helpful remarks,
to Arkadiy~Onishchik for attracting author's
attention to the used algebraic technique,
and to Ruud~Martini and Vladimir~Sokolov 
for useful discussions.

%%%%%%%%%%%%%%%%%%%%%%%%%%%%%%%%%%%%%%%%%%%%%%%%%%


\begin{thebibliography}{99}

\bibitem{rb}
A.~V.~Bocharov, V.~N.~Chetverikov, S.~V.~Duzhin, N.~G.~Khor{\cprime}kova,
I.~S.~Krasil{\cprime}shchik, A.~V.~Samokhin, Yu.~N.~Torkhov, 
A.~M.~Verbovetsky, and  A.~M.~Vinogradov.
\emph{Symmetries and Conservation Laws for Differential Equations of
Mathematical Physics}.
Edited and with a preface by Krasil\cprime shchik and Vinogradov.
Amer. Math. Soc., Providence, RI, 1999.

\bibitem{grif}
R.~L.~Bryant and P.~A.~Griffiths.
Characteristic cohomology of differential systems. {I}.
{G}eneral theory. \emph{J. Amer. Math. Soc.} \textbf{8}
(1995), 507--596.

\bibitem{foltinek}
K.~Foltinek.
Conservation laws of evolution equations: generic non-existence. 
\emph{J. Math. Anal. Appl.} \textbf{235} (1999), 356--379.


\bibitem{kurosh}
A.~Kurosh. 
\emph{Higher algebra}. Mir Publishers, Moscow, 1972.
     
\bibitem{larkin}
N.~Larkin. On one problem of transonic gas dynamics. 
\emph{Mat. Contemp.} \textbf{15} (1998), 169--186.

\bibitem{mikh}
A.~V.~Mikhailov, A.~B.~Shabat and R.~I.~Yamilov.
A symmetric approach to the classification of nonlinear
equations. {C}omplete lists of integrable systems.
\emph{Russian Math. Surveys} \textbf{42} (1987), 1--63.

\bibitem{vis}
J.~Mukherjee and A.~Roy~Choudhury.
On the Lie symmetry, conservation laws and 
Painlev\'e analysis for the viscous transonic equation. 
\emph{Ann. Physik (7)} \textbf{46} (1989), 6--92. 

\bibitem{olver}
P.~J.~Olver.
\emph{Applications of Lie Groups to Differential Equations}.
Springer-Verlag, New York, 1993.

\bibitem{ryzhov}
O.~S.~Ryzhov.
Viscous transonic flows.
\emph{Annual review of fluid mechanics, Vol. 10}, 65--92.
Annual Reviews, Palo Alto, Calif., 1978.


\bibitem{pdes}
M.~E.~Taylor. 
\emph{Partial differential equations. III. Nonlinear equations.}
Springer-Verlag, New York, 1996.

\bibitem{vel}
P.~A.~Vel\cprime misov, S.~V.~Fal\cprime kovi\v c. 
On the theory of transonic flows of a viscous gas. (Russian) 
\emph{Izv. Vys\v s. U\v cebn. Zaved. Matematika} {\bf 1974}, no.~5, 52--61.

\bibitem{vin}
A.~M. Vinogradov.
The $\mathcal{C}$-spectral sequence, Lagrangian formalism, and
conservation laws.
I.\ The linear theory. II.\ The nonlinear theory.
\emph{J. Math. Anal. Appl.} \textbf{100} (1984), 1--129.

\end{thebibliography}
\end{document}